\documentclass[a4paper,12pt]{article}
\usepackage{amssymb,amsfonts,amsmath,float}

\title{On the entropy gain under the action of amplitude damping channel on qutrit}

\author{Grigori G. Amosov, Andrey I. Dnestryan\\
Steklov Mathematical Institute\\ Moscow Institute of Physics and Technology}

\begin{document}

\maketitle

\begin{abstract}
After realising qutrit in the form of bipartite system we estimate from below the entropy gain under the action
of the amplitude damping channel.
 \end{abstract}

\section {Introduction}

Recently the method of the qubit portrait for qudit states was introduced \cite {Chernega}. 
Let us consider a qutrit state
$$
\rho =\left (\begin{array}{ccc}\rho _{11}&\rho _{12}&\rho _{13}\\ \rho _{21}&\rho _{22}&\rho _{23}\\ \rho _{31}&\rho _{32}&\rho _{33}\end{array}\right ).
$$
Then, one of its possible qubit portrait can be defined as
$$
\sigma =\left (\begin{array}{cc}\rho _{11}+\rho _{22}&\rho _{13}\\ \rho _{31}&\rho _{33}\end{array}\right ). 
$$
In particular, this method
allows to estimate the von Neumann entropy $S(\rho )=-Tr\rho\log\rho$ of a qutrit state $\rho $ by means of the entropies of its qubit portraits \cite {Manko}.
On the other hand, for bipartite systems there is a possibility to estimate the entropy gain under the action of quantum channels
having a special form \cite {Amosov}. Suppose that a state $\rho \in \mathfrak {S}(\mathcal {H}\otimes \mathcal {K})$ 
has the form
\begin{equation}\label{hr}
\rho =\sum \limits _{jk}\alpha _{jk}|e_j\rangle\langle e_k|\otimes |h_j\rangle\langle h_k|,
\end{equation}
where $(e_j)$ is an orthonormal basis in $H$, $(h_j)$ are unit vectors in $K$ and $(\alpha _{jk})$ is a positive definite matrix
with unit trace.
Given a quantum channel $\Omega :\mathfrak {S}(\mathcal {K})\to \mathfrak {S}(\mathcal {K})$ it is possible to obtain the estimation from below for
the entropy gain under the action of the channel $Id\otimes \Omega $ on a state of the form (\ref {hr}) 
as follows \cite{Amosov}
\begin{equation}\label{gain}
S((Id\otimes \Omega )(\rho))-S(\rho)\ge \sum \limits _j\alpha _{jj}S(\Omega (|h_j\rangle\langle h_j|). 
\end{equation}
In the present paper we shall include a qutrit state into the bipartite system and
estimate the entropy gain under the action of the quantum channel being a tensor product of the amplitude damping channel
and the identity channel. The right hand part of our estimation will consist of the entropy of some qubit state.

\section{Decomposing density matrix into the sum of tensor products}

Consider a density operator for some two-mode state:
\begin{equation}\label{1}
\rho=\begin{pmatrix}
\rho_{11} & \rho_{12} & \rho_{13} & 0 \\
\rho_{21} & \rho_{22} & \rho_{23} & 0 \\  
\rho_{31} & \rho_{32} & \rho_{33} & 0 \\  
0 & 0 & 0 & 0 \\
\end{pmatrix}
\end{equation}
The question arises: when $\rho $ can be represented as a sum of the form
\begin{equation}\label{2}
\rho=\sum_{n,m}\alpha_{nm}|e_n\rangle\langle e_m| \otimes |h_n\rangle\langle h_m|,
\end{equation}
where $|e_n\rangle$ is a orthonormal basis in $\mathcal{H}$, f.e. $|e_1\rangle={1 \choose 0}$,
 $|e_2\rangle={0 \choose 1}$, $|h_n\rangle$ are unit (non collinear in general) vectors $\mathcal{K}$ and $(\alpha_{nm})$ 
 is a positive definite matrix with unit trace. Vectors $|h_n\rangle$ are complex but without loss of generality we consider only the second coordinate to be complex in each of them. 
 We denote $|h_1\rangle={x \choose y}$ and $|h_2\rangle={a \choose b}$, $x^2+|y|^2=a^2+|b|^2=1$.
 So we have:
$$\rho=\begin{pmatrix}
\alpha_{11}\begin{pmatrix}x^2 & x\overline y\\xy & |y|^2 \end{pmatrix}& \alpha_{12}\begin{pmatrix}ax & x \overline b\\ay & y\overline b \end{pmatrix} \\
\alpha_{21}\begin{pmatrix}ax & a\overline y\\bx & b \overline y \end{pmatrix} & \alpha_{22}\begin{pmatrix}a^2 & a\overline b\\ab & |b|^2 \end{pmatrix} 
\end{pmatrix}$$
Comparing with the original expression for $ \rho $ we do the obvious conclusion that $ b = 0 $ and $ a = 1 $. Further, since $ (\alpha_ {nm}) $ is a positive definite matrix, then $ \alpha_ {11}, \; \alpha_ {22} \in \mathbb {R} $. Consequently, the matrix $ \rho $ is of the form
$$\rho=
\begin{pmatrix}\alpha_{11}x^2 & \alpha_{11}x\overline y & \alpha_{12}x & 0 \\
\alpha_{11}xy & \alpha_{11}y\overline y & \alpha_{12}y & 0  \\
\alpha_{21}x & \alpha_{21}\overline y & \alpha_{22} & 0 \\  0 & 0 & 0 & 0
\end{pmatrix}$$
Matrix $\rho$ is Hermitian ($\rho_{ij}=\overline{\rho_{ji}}$), so $\alpha_{12}=\overline{\alpha_{21}} \in \mathbb{C}$. We note that the second and the third minors of matrix are equal to zero, therefore 
\begin{equation}\label{minor}
\rho_{11}\rho_{22}=\rho_{12}\rho_{21} 
\end{equation}
and
$$\begin{vmatrix}
\rho_{11} & \rho_{12} & \rho_{13}  \\
\rho_{21} & \rho_{22} & \rho_{23}  \\  
\rho_{31} & \rho_{32} & \rho_{33}
\end{vmatrix}=0$$ 
Thus not any density matrix can be decomposed into (\ref {2}). We got some conditions on the density matrix $ \rho $ which allows us to decompose it into a sum of tensor products (\ref {2}).

\section{Entropy gain}
\par
It is known that if $\rho$ has the form (\ref {2}), then for any quantum channel $\Omega: \sigma (\mathcal{K}) \to \sigma (\mathcal{K})$ Estimation (\ref {gain}) is valid.
Consider the amplitude damping channel
$$\Omega(\rho)=\Omega\begin{pmatrix}\rho_{11} & \rho_{12}\\\rho_{21} & \rho_{22} \end{pmatrix}=\begin{pmatrix}\rho_{11}+\gamma\rho_{22} & \sqrt{1-\gamma}\rho_{12}\\ \sqrt{1-\gamma}\rho_{21} & (1-\gamma)\rho_{22} \end{pmatrix}$$
Amplitude damping channel describes energy dissipation from a quantum system such as spontaneous emission of an atom and relaxation of a spin
system at high temperature into the equilibrium state \cite {Neilson}. It is a non-unitary channel with Kraus operators
$$A_0=\begin{pmatrix}1 & 0\\ 0 & \sqrt{1-\gamma} \end{pmatrix}, \quad A_1=\begin{pmatrix}0 & \sqrt{\gamma}\\ 0 & 0 \end{pmatrix}$$
Using (\ref {gain}) we obtain an estimation of the entropy gain for this channel of the form
$$-Tr\begin{pmatrix}
\rho_{11}+\gamma\rho_{22} & \sqrt{1-\gamma}\rho_{12} & \rho_{13} \\
\sqrt{1-\gamma}\rho_{21} & (1-\gamma)\rho_{22} & \sqrt{1-\gamma}\rho_{23} \\  
\rho_{31} & \sqrt{1-\gamma}\rho_{32} & \rho_{33}\\
\end{pmatrix}\ln \begin{pmatrix}
\rho_{11}+\gamma\rho_{22} & \sqrt{1-\gamma}\rho_{12} & \rho_{13} \\
\sqrt{1-\gamma}\rho_{21} & (1-\gamma)\rho_{22} & \sqrt{1-\gamma}\rho_{23} \\  
\rho_{31} & \sqrt{1-\gamma}\rho_{32} & \rho_{33} \\  
\end{pmatrix} +$$
$$Tr\begin{pmatrix}
\rho_{11} & \rho_{12} & \rho_{13} \\
\rho_{21} & \rho_{22} & \rho_{23}  \\  
\rho_{31} & \rho_{32} & \rho_{33} \\  
\end{pmatrix}\ln\begin{pmatrix}
\rho_{11} & \rho_{12} & \rho_{13} \\
\rho_{21} & \rho_{22} & \rho_{23} \\  
\rho_{31} & \rho_{32} & \rho_{33} \\  
\end{pmatrix} \ge $$
$$-(\rho _{11}+\rho _{22})Tr\begin{pmatrix}
\frac {\rho_{11}+\gamma\rho_{22}}{\rho _{11}+\rho _{22}} & \frac {\sqrt{1-\gamma}\rho_{12}}{\rho _{11}+\rho _{22}}\\ \frac {\sqrt{1-\gamma}\rho_{21}}{\rho _{11}+\rho _{22}} & \frac {(1-\gamma)\rho_{22}}{\rho _{11}+\rho _{22}}
\end{pmatrix}\ln\begin{pmatrix}
\frac {\rho_{11}+\gamma\rho_{22}}{\rho _{11}+\rho _{22}} & \frac {\sqrt{1-\gamma}\rho_{12}}{\rho _{11}+\rho _{22}}\\ \frac {\sqrt{1-\gamma}\rho_{21}}{\rho _{11}+\rho _{22}} & \frac {(1-\gamma)\rho_{22}}{\rho _{11}+\rho _{22}}
\end{pmatrix}$$
Note that if $\gamma =0$ we get zeros in both the parts of the estimation due to the state
$$
\begin{pmatrix}
\frac {\rho_{11}}{\rho _{11}+\rho _{22}} & \frac {\rho_{12}}{\rho _{11}+\rho _{22}}\\ \frac {\rho_{21}}{\rho _{11}+\rho _{22}} & \frac {\rho_{22}}{\rho _{11}+\rho _{22}}
\end{pmatrix}
$$
is pure due to (\ref {minor}).

\section{Conclusion}

To estimate the entropy gain under the action of quantum channel we have embedded qutrit into the bipartite system. Then, for the case of amplitude damping channel
we have applied the known result of \cite {Amosov}. The right hand part of our estimation includes the von Neumann entropy of a qubit state.

\section{Acknowledgments}

The work of the first author is partially supported by RFBR grant 12-01-00319 and by the RAS research program.

\end{document}